\def\year{2022}\relax
\documentclass[letterpaper]{article} 
\usepackage{aaai22}  
\usepackage{times}  
\usepackage{helvet}  
\usepackage{courier}  
\usepackage[hyphens]{url}  
\usepackage{graphicx} 
\usepackage{natbib}  
\usepackage{caption} 
\DeclareCaptionStyle{ruled}{labelfont=normalfont,labelsep=colon,strut=off} 
\usepackage{algorithm}
\usepackage{algorithmic}

\usepackage{newfloat}
\usepackage{listings}
\floatstyle{ruled}
\newfloat{listing}{tb}{lst}{}
\floatname{listing}{Listing}

\usepackage{booktabs}       
\usepackage{amsfonts}       
\usepackage{nicefrac}       
\usepackage{microtype}      
\usepackage{xcolor}         

\usepackage{epsfig}
\usepackage{amsmath}
\usepackage{amssymb}
\usepackage{color, colortbl}
\usepackage{pifont}
\usepackage{multirow}
\usepackage{adjustbox}
\usepackage{enumerate}
\usepackage{xspace}
\usepackage{subcaption}

\definecolor{Gray}{gray}{0.9}
\definecolor{LightCyan}{rgb}{0.88,1,1}
\definecolor{Yellow}{rgb}{1,1,0.88}

\makeatletter
\DeclareRobustCommand\onedot{\futurelet\@let@token\@onedot}
\def\@onedot{\ifx\@let@token.\else.\null\fi\xspace}

\def\eg{\emph{e.g}\onedot} 
\def\ie{\emph{i.e}\onedot}

\makeatother

\title{Model-Based Image Signal Processors via Learnable Dictionaries}
\author {
    Marcos V.~Conde, Steven McDonagh, Matteo Maggioni, Aleš Leonardis, Eduardo P\'erez-Pellitero
}
\affiliations {
    Huawei Noah's Ark Lab
}

\begin{document}

\maketitle

\begin{abstract}
Digital cameras transform sensor RAW readings into RGB 
images by means of their Image Signal Processor (ISP). 
Computational photography tasks such as image denoising and colour constancy are commonly performed in the RAW domain, in part due to the inherent hardware design, but also due to the appealing simplicity of noise statistics that result from the direct sensor readings. 
Despite this, the availability of RAW images is limited in comparison with the 
abundance and diversity of available RGB data. Recent approaches have attempted to bridge this gap by estimating the RGB to RAW mapping: handcrafted model-based methods that are interpretable and controllable usually require manual parameter fine-tuning, while end-to-end learnable neural networks require large amounts of training data, at times with complex training procedures, and generally lack interpretability and parametric control.
Towards addressing these existing limitations,
we present a novel hybrid model-based and data-driven ISP that builds on canonical ISP operations and is both learnable and interpretable.
Our proposed invertible model, capable of bidirectional mapping between RAW and RGB domains, employs 
end-to-end learning of rich parameter representations, \ie dictionaries
, that are free from direct parametric supervision and additionally enable simple and plausible data augmentation.
We evidence the value of our data generation process by extensive experiments under both RAW image reconstruction and RAW image denoising tasks, obtaining state-of-the-art performance in both. 
Additionally, we show that our ISP can learn meaningful mappings from few data samples, and that denoising models trained with our dictionary-based data augmentation are competitive despite having only few or zero ground-truth labels. 
\end{abstract}


\section{Introduction}
\label{sec:intro}

Advances in Convolutional Neural Networks have made great strides in many computer vision applications in the last decade, in part thanks to the proliferation of camera devices and the resulting availability of large-scale image datasets. The majority of these datasets contain sRGB image data, which is obtained via an in-camera Image Signal Processor (ISP) that converts the camera sensor's RAW readings into perceptually pleasant RGB images, suitable for the human visual system. 
However, the characteristics of RAW imagery (\eg linear relationship with scene irradiance, raw and untampered signal and noise samples) are often better suited for the ill-posed, inverse problems that commonly arise in low-level vision tasks such as denoising, demosaicing, HDR, super-resolution~\cite{qian2019trinity, Abdelhamed_2018_CVPR, Wronski2019hanheldsuper, Gharbi2016-deepjoint, Liu_2020_CVPR-hdr-isp}. For tasks within the ISP, this does not come as a choice but rather a must, as the input domain is necessarily in the RAW domain due to the camera hardware design~\cite{buckler2017reconfiguringisp, ignatov2021realtime}. 

Unfortunately, RAW image datasets are not nearly as abundant and diverse as their RGB counterparts, and thus some of the performance potential of CNN-based approaches cannot be fully utilized. To bridge this gap, recent methods aim to estimate the mapping from sRGB.

The recent work of \citeauthor{Brooks_2019_CVPR} introduces a generic camera ISP model composed of five canonical steps, each of them approximated by an invertible, differentiable function. Their proposed ISP can be conveniently plugged-in to any RGB training pipeline to enable RAW image processing. As each of the functions is constrained to perform a single task within the ISP, intermediate representations are fully interpretable, allowing for complete flexibility and interpretability in the ISP layout. Despite successful application to image denoising, this approach requires manually setting the true internal camera parameters, and these cannot be learnt from data. Although some DSLR cameras do provide access to such parameters, ISP layouts and their related inner settings are generally protected and inaccessible to the end user.

Alternative recent learning-based approaches ~\cite{Punnappurath2020-raw-recons, Zamir_2020_CVPR, xing21invertible} attempt to learn the ISP in an end-to-end manner, in order to circumvent the noted problems. Focused on the RAW data synthesis from sRGB images, CycleISP~\cite{Zamir_2020_CVPR} adopts \emph{cycle consistency} to learn both the forward (RAW-to-RGB) and reverse (RGB-to-RAW) directions of the ISP using 2 different networks and is trainable end-to-end. The authors show that RGB data can then be leveraged successfully to aid a RAW denoising task. The ISP is thus learned as a black-box, is not modular and therefore lacks both 
interpretability for intermediate representations and control of the ISP layout. However, these traits can be considered 
important when training for specific intermediate tasks within the ISP (\eg colour constancy). Additionally, as there is no model regularization, training CycleISP remains a complex procedure, requiring large amounts of RAW data.

Contemporary to our presented work, the InvISP~\cite{xing21invertible} proposes the camera ISP model as an invertible ISP using a single invertible neural network \cite{Kingma-NEURIPS2018-flow, ho2019pmlr-flow+} to perform both the RAW-to-RGB and RGB-to-RAW mapping. This normalizing-flow-based approach has the advantages of being invertible and learnable, however, as CycleISP lacks of interpretability and control, requires large amounts of training data and its constrained by the invertible blocks (\ie input and output size must be identical).

In this paper we introduce a novel hybrid approach that tackles the aforementioned limitations of ISP modelling and retains the best of both model-based and end-to-end learnable approaches~\cite{shlezinger2021modelbased}. We propose a modular, parametric, model-driven approach with a novel parameter dictionary learning strategy that builds on~\citeauthor{Brooks_2019_CVPR} We further improve this flexible, interpretable and constrained ISP architecture with additional lens shading modelling and a more flexible parametric tone mapping. To address the lack of in-camera parameters, discussed previously, we design an end-to-end learnable dictionary representation of inner camera parameters. This provides a set of parameter basis for optimal end-to-end reconstruction, and enables unlimited data augmentation in the RAW image manifold. Our proposed method is modular, interpretable and is governed by well-understood camera parameters. It provides a framework to learn an end-to-end ISP and related parameters from data. Moreover, 
it can be learnt successfully from 
very few samples, even when corrupted by noise.
Note that we focus on the RAW reconstruction task and its downstream applications (\eg denoising, HDR imaging). The forward pass or RAW-to-RGB processing, despite related, is a different research problem~\cite{Ignatov_2020_CVPR_Workshops, Schwartz_2019-deepisp, Ling2021TIP-cameranet}.

\textbf{Our main contributions} can be summarized as follows: (1) 
a modular and differentiable ISP model, composed of canonical 
camera operations and 
governed by interpretable 
parameters, (2) 
a training mechanism that, in conjunction with our model contribution, is capable of end-to-end learning of 
rich parameter representations, \ie 
dictionaries or basis and related linear decomposition decoders, 
that result in compact ISP models, 
free from direct parameter supervision, (3)~extensive experimental investigation; 
our learned RGB-to-RAW mappings are used to enable data augmentation towards down-stream task performance improvement, in multiple data regimes of varying size and noise.

\section{Image Signal Processor}
\label{sec:isp}
The group of operations necessary to convert the camera sensor readings into natural-looking RGB images are generally referred to as the Image Signal Processor (ISP). There is great variability in ISP designs with varying levels of complexity and functionalities, however a majority of them contain at least a number of operations that are generally considered to be a canonical representation of a basic ISP, namely white balance, color correction, gamma expansion and tone mapping~\cite{msbrown2016CVPR-Tutorial, Heide2014acm-flexisp, delbracio2021mobile}. \citeauthor{Brooks_2019_CVPR} introduces a modular, differentiable ISP model where each module is an invertible function that approximates the aforementioned canonical operations. In this section we review that work, and introduce notation and parameter details about each operation as well as the complete function composition.

Let us initially define two images spaces: the RAW image domain $\mathcal{Y}$ and the sRGB image domain $\mathcal{X}$. The transformation done by the camera ISP can thus be defined as $ f : \mathcal{Y} \rightarrow \mathcal{X}$.  Intuitively, we can define a modular ISP function $f$ as a composite function as follows:
\begin{equation}
x=(f_{n}\circ\cdots\circ f_{2}\circ f_{1})(y,p_{n},\ldots,p_{2}, p_{1}),
\end{equation}
where $f_i$ is a function with related parameters $p_i$ for a composition of arbitrary length $n$. 
In order to recover a RAW image $y$ from the respective sRGB observation $x$ (\ie a mapping from $\mathcal{X} \rightarrow \mathcal{Y}$) we can choose ${f_i}$ to be invertible and tractable bijective functions:
\begin{equation}
y=(f_{1}^{-1}\circ f^{-1}_{2} \circ\cdots\circ f^{-1}_{n})(x,p_{1},p_{2},\ldots,p_{n}).
\end{equation}

\subsection{Colour Filter Array Mosaic}
\label{sec:isp:cfa}
Camera sensors use a Colour Filter Array (CFA) in order to capture wavelength-specific colour information. The sensor is covered with a pattern of colour filters arranged in a given spatial pattern, \eg the well known Bayer pattern, which is a $2\times2$ distribution of R - G - G - B colours, that effectively produces a single colour measurement per spatial position. 

In order to obtain the missing colour samples for each spatial position, the so called demosaicing methods aim to recover the missing pixels, commonly an ill-posed problem which, for the sake of simplicity, we will address as a simple bilinear interpolation: $f_6(y) = \textrm{bic}(y)$. The inverse of this function, is however, a straightforward mosaicing operation. It can be defined as:
\begin{equation}
    f_6^{-1}(x_5, k_m) = x_5*k_m,
\end{equation} where $*$ denotes a convolution with kernel $k_m$ containing the mosaicing pattern, generally strictly formed by $\{0,1\}$.

\subsection{Lens Shading Effect}
\label{sec:isp:lens}
Lens Shading Effect  (LSE) is the phenomenon of the reduced amount of light captured by the photoreceptor when moving from the center of the optical axis towards the borders of the sensor, mostly caused by obstruction of elements involved in the lens assembly. We can define this function as:
\begin{equation}
    f_5 (x_5, M) = x_5 \odot M,
\end{equation}
where $M$ is a mask containing per-pixel lens-shading gains. This can be inverted by 
inverting each per-pixel gain.

\subsection{White Balance and Digital Gain}
\label{sec:isp:wb}
The White Balance (WB) stage aims to neutralize the scene light source colour such that after correction its appearance matches that of an achromatic light source. In practice, this is achieved by a global per-channel gain for two of the colours, \ie red and blue gains, namely $g_r$ and $g_b$ respectively, which we arrange in a three colour vector 
$g_\textrm{wb} = \left[\begin{array}{ccc}
g_{r} & 1 & g_{b}\end{array}\right]$. 
This scene illuminant is generally estimated heuristically, although more sophisticated approaches have also been explored ~\cite{gijsenij2009perceptual,barron2017fast,hernandez2020multi}.

WB is normally applied in conjunction with a scalar digital gain $g_d$, which is applied globally to all three channels, and scales the image intensities as desired. This process can be conveniently described as a convolution: 
\begin{equation}
    f_4 (x_4, g_{wb}, g) = x_4*(g_d~g_{\textrm{wb}}).
\end{equation}
To obtain the inverse function $f_4^{-1}$, we just invert each of the gains individually, but instead of using the naive division $\nicefrac{1}{g}$, we follow the 
highlight-preserving cubic transformation of~\citeauthor{Brooks_2019_CVPR}

\subsection{Color Correction}
\label{sec:isp:cc}
The ISP converts the sensor color space into the output color space. This step is often necessary as the CFA colour spectral sensitivity does not necessarily match the output color standard, \eg sRGB~\cite{msbrown2016CVPR-Tutorial, Afifi2021TPAMI-CIEXYZNet-raw-recons}. A global change in the color space can be achieved with a $3\times3$ Color Correction Matrix (CCM):
\begin{equation}
    f_3 (x_3, \mathbf{C_m}) = \mathbf{X}_{3}\mathbf{C}_{m},
\end{equation}
where $\mathbf{C}_{m}$ denotes a CCM parameter and $\mathbf{X}_{3}$ denotes $x_3$ reshaped for convenience as a matrix, \ie $\mathbf{X}_{3} \in \mathbb{R}^{hw\times3}$. Similarly to $f_3$, we can obtain $f_3^{-1}$ by using $\mathbf{C}_m$ pseudo-inverse.

\subsection{Gamma Correction}
The camera sensor readings are linearly proportional to the light received, however the human visual system does not naturally perceive light linearly, but rather is more sensitive to darker regions. Thus it is common practice to adapt the linear sensor readings with a gamma logarithmic function:
\begin{equation}
    f_2(x_2, \gamma)=\max(x_2, \epsilon)^{1/\gamma},
\end{equation}
where $\gamma$ is a parameter regulating the amount of compression/expansion, generally with values around $\gamma=2.2$. The inverse function can be defined as follows: 
\begin{equation}
    f_2^{-1}(x_1, \gamma)=\max(x_1, \epsilon)^{\gamma}.
\end{equation}

\subsection{Tone Mapping}
\label{sec:isp:tone}
Tone Mapping Operators (TMOs) have been generally used to adapt images to their final display device, the most common case being the TMO applied to High Dynamic Range Images (HDRI) on typical Low Dynamic Range display devices. As opposed to using an S-shaped polynomial function as proposed by \cite{Reinhard2002-tonemap,Brooks_2019_CVPR}, we can use instead a parametric piece-wise linear function that we model as a shallow convolutional neural network~\cite{Punnappurath2020-raw-recons} composed only by $1\times1$ kernels and ReLU activations:
\begin{equation}
    f_1(x_1, \theta_f) = \phi_t(x_1, \theta_f),
\end{equation}
where $\phi$ is a shallow CNN with learnable parameters $\theta_f$ for the forward pass. A different set of weights $\theta_r$ can be optimized for the reverse pass.

\section{Learning Parameter Representations}
\label{sec:our-method}

\begin{figure*}
   \begin{center}
   \includegraphics[width=0.89\textwidth]{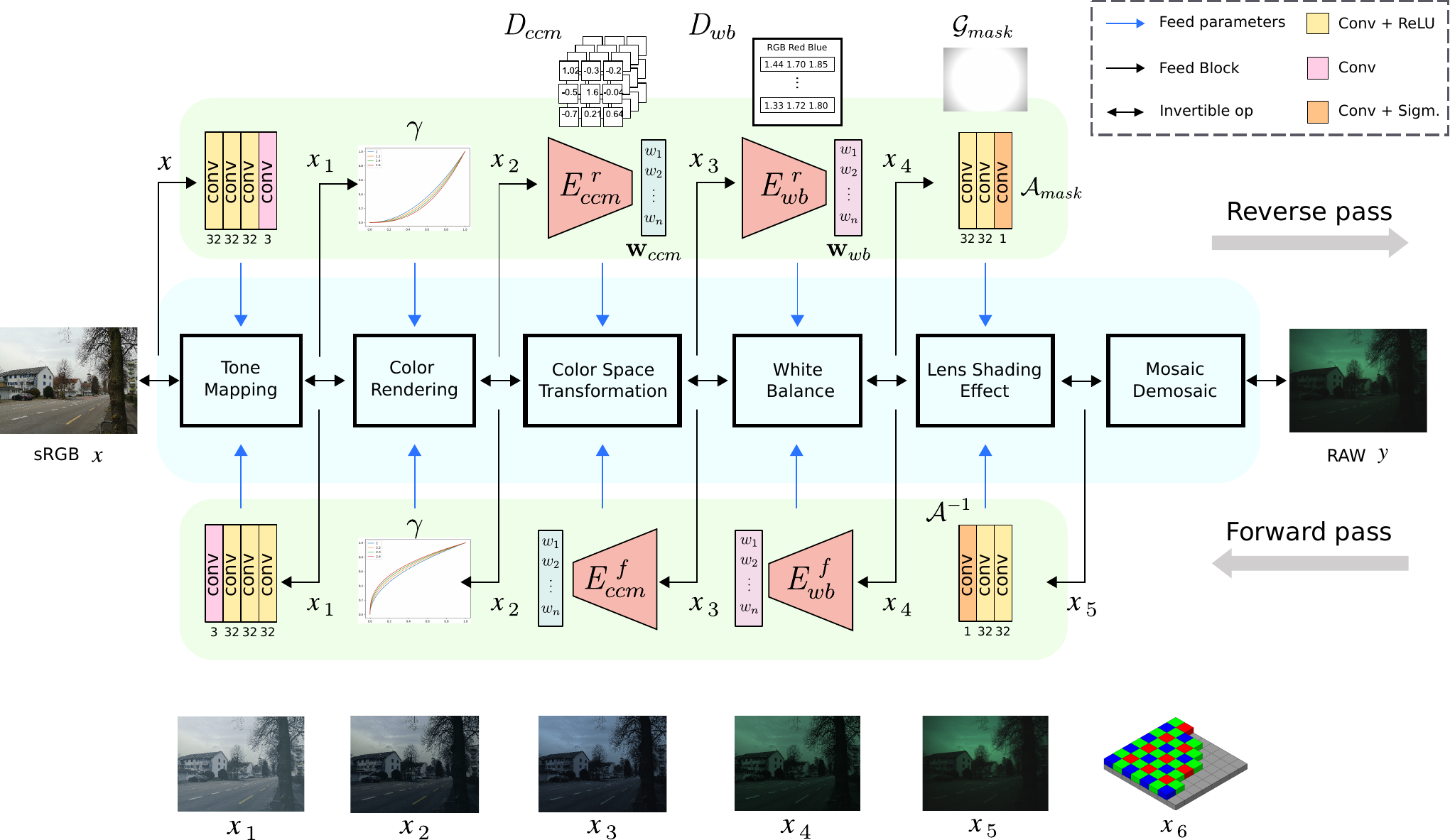}
   \end{center}
   \caption{A visualization of our proposed model using as backbone (blue) the classical ISP operations described in Section \ref{sec:isp}, and additional learning component (green) described in Section \ref{sec:our-method}.
   For visualization purposes, RAW images are visualized through bilinear demosaicing. This figure is best viewed in the electronic version.
   }
   \label{fig:main_isp}
\end{figure*}

In the previous section we introduced a modular and parametric ISP model, however that model alone does not allow for end-to-end training. In this section we introduce a strategy to enable end-to-end training for the presented ISP model. To the best of our knowledge, our method is the first data-driven, model-based approach to tackle the reverse ISP problem.

In Figure \ref{fig:main_isp} we show an overview of our proposed approach, formed by a 6-stage ISP model (in blue colour) and separate networks that learn parameter dictionaries and feed parameters to the model (in green colour).


\subsection{Parameter Dictionaries}
\label{dictionaries}

\begin{figure}
   \begin{center}
   \includegraphics[width=\linewidth]{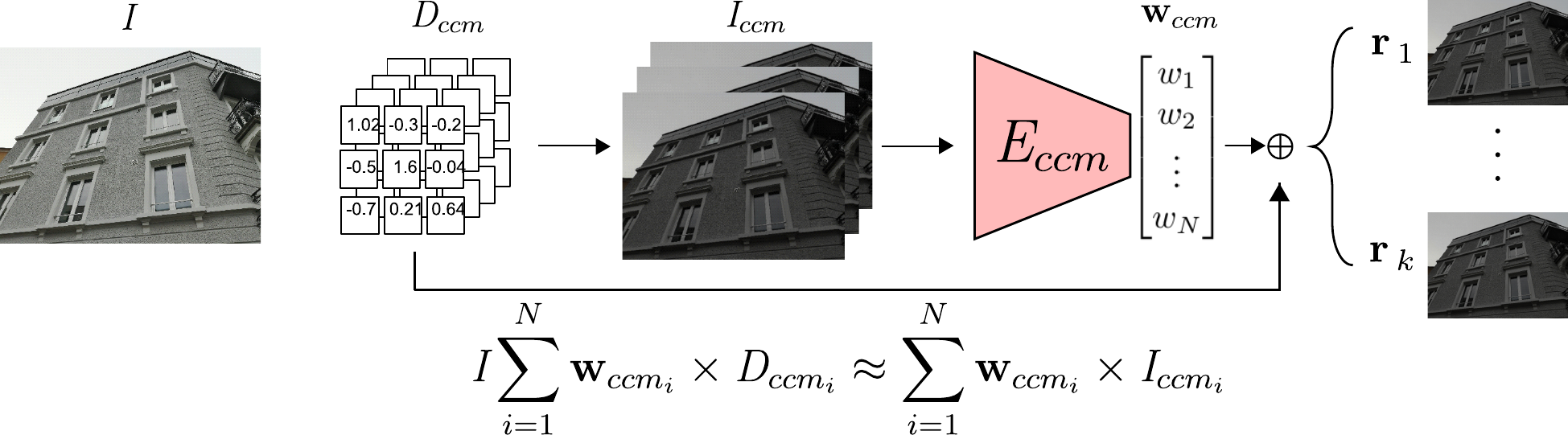}
   \end{center}
   \caption{Dictionary-based Color Space Transformation. During training, $\textit{D}_{ccm}$ and $\textit{E}_{ccm}$ are learned together and generate a single output. At inference time, dictionary augmentations are used to generate $k$ samples from a single RGB input. The $\mathbf{r}_k$ represent the random perturbations added to $\mathbf{w}_{ccm}$.}
   \label{fig:dicts}
\end{figure}

\textbf{Color Correction} Modern smartphone cameras typically use different CCMs depending on specific light conditions and capture modes, so any method that assumes a single CCM mode might struggle to cope with colour variability. Additionally, an ISP model might be trained to reconstruct RAW images captured with different cameras and thus also different ISP and CCMs. As previously discussed, these matrices are generally not accessible to the end user.
In order to learn the color space transformation done by the ISP, we create a dictionary $\textit{D}_{ccm} \in \mathbb{R}^{N \times 3 \times 3}$ of size $N$, where each atom is a CCM. To preserve the significance and physical meaning of these matrices, and avoid learning non-realistic parameters, we constrain the learnt atoms in the dictionary by column-normalizing each matrix following the $\ell_1$ norm, as this is one of the most representative properties of realistic CCMs~\cite{Brooks_2019_CVPR, Koskinen2019-reverse-isp-icip}.
We perform the color correction as a convolution operation, where the convolutional kernels are the atoms of $\textit{D}_{ccm}$ and the input is the intermediate representation from the previous function in the ISP model. As the result of this operation we obtain $\textit{I}_{ccm} \in \mathbb{R}^{N \times H \times W \times 3}$, which represents $N$ RGB images, each one the result of applying each atom to the input image.
This representation $\textit{I}_{ccm}$ passes through a CNN encoder $\textit{E}_{ccm}$ that produces
a vector of weights $\mathbf{w}_{ccm} \in \mathbb{R}^{N}$. 
The resultant color transformed sRGB image is obtained as a linear combination of $\textit{I}_{ccm}$ and $\mathbf{w}_{ccm}$, which is equivalent to linearly combining the atoms in the dictionary, and applying the resultant CCM to the image.
As illustrated in Figure \ref{fig:dicts}, the model simultaneously learns $\textit{D}_{ccm}$ and $\textit{E}_{ccm}$. This novel dictionary representation of the camera parameters can allow learning the CCMs of various cameras at once.
Note that the encoder $\textit{E}^{r}_{ccm}$ used during the reverse pass is different from the $\textit{E}^{f}_{ccm}$ used in the forward pass as we show in Figure~\ref{fig:main_isp}, however, both encoders have the same functionality.\\
%
\textbf{Digital Gain and White Balance}
Similarly to the CCM dictionaries, we define $\textit{D}_{wb} \in \mathbb{R}^{N \times 3} $ as a dictionary of $N$ white balance and digital gains, thus, each atom is a triplet of scalars $(g_d~g_{r}~g_{b})$. 
We apply each atom $g$ from the dictionary as described by~\citeauthor{Brooks_2019_CVPR} and obtain $\textit{I}_{wb} \in \mathbb{R}^{N \times H \times W \times 3}$, which represents a linear decomposition of the results from applying each $g_i$ to the input image. 
An encoder $\textit{E}_{wb}$ produces a set of weights $\mathbf{w}_{wb} \in \mathbb{R}^{N}$ from such representation. Note that this encoder is different from the $\textit{E}_{ccm}$ used in the color correction step.
The encoder and dictionary are learned jointly in the optimization. The linear combination of $\textit{I}_{wb}$ and $\mathbf{w}_{wb}$ produce our white balanced image.
To ensure we keep the physical meaning and interpretability of the learnt WB gains, we found sufficient to initialize the atoms in $\textit{D}_{wb}$ using a uniform distribution $\mathcal{U}(1,2)$ that encourages appropriate behaviour on the learnt gain vectors: non-negative and a reasonable range for pixel gains (i.e. approximately $[1,2]$)~\cite{Brooks_2019_CVPR}.
As we show in Figure~\ref{fig:main_isp}, the reverse pass encoder  $\textit{E}^{r}_{wb}$ is different from the forward pass encoder $\textit{E}^{f}_{wb}$, however, both work in the same way.\\
\textbf{Dictionary Augmentations} Two learnt dictionaries, \ie $\textit{D}_{ccm}$ and $\textit{D}_{wb}$, can be interpreted as a set of basis describing the parameter space. For a given RGB input, encoders find the combination of atoms in the dictionary that optimize the RAW image reconstruction, represented as a vector of weights $\mathbf{w}$.
We can further exploit this representation to generate unlimited RAW images by adding small perturbations to the optimal weights $\mathbf{w}_{ccm}$ and $\mathbf{w}_{wb}$, by \eg adding Gaussian noise. These dictionaries represent a convex hull of plausible parameters, and thus any variation within that space is likely to result in useful data for downstream tasks. Figure~\ref{fig:dicts} shows this process. Once the dictionaries are learnt, it is possible to remove the related encoders (0.5M parameters) and sample convex combinations of the elements in the dictionary, reducing considerable the complexity of our model, and allowing to process high resolution images with low-latency.

\subsection{Piecewise Linear Tone Mapping}
\label{tonemap}

Tone mapping~\cite{mantiuk2009color} is a technique used by the camera to map one set of colors to adjust the image’s aesthetic appeal by compressing the high-intensity and low-intensity values more than mid-intensity values. 
A tone map is usually designed as a 1D lookup table (LUT) that is applied per color channel to adjust the tonal values of the image, or as a “smoothstep” S-curve. To reconstruct RAW data tones from sRGB is challenging, especially at the over-exposed regions and high-dynamic range images require more sophisticated tone mapping. We propose a piecewise linear CNN as a learnable tone map~\cite{Punnappurath2020-raw-recons}.
In the forward pass, tone mapping is performed using ${f}_{1}$. At the reverse pass, we perform the inverse tone mapping using ${f}^{-1}$. Both functions are shallow CNNs implemented using pixel-wise convolutional blocks to constraint the possible transformations and easily control the network, and representing by its definition piecewise linear models.

\subsection{Lens Shading Modelling}
\label{lse}

Due to sensor optics, the amount of light hitting the sensor falls off radially towards the edges and produces a vignetting effect; 
known as lens shading. 
A typically early ISP stage constitutes Lens Shading Correction (LSC)~\cite{young2000shading} and is used to correct the effects of uneven light hitting the sensor, towards providing a uniform light response. This is done by applying a mask, typically pre-calibrated by the manufacturer, to correct for 
non-uniform light fallout effects~\cite{delbracio2021mobile}. 
Modelling of the ISP therefore requires 
a method to add or correct the Lens Shading Effect (LSE) by modelling such a mask. We propose to model this mask 
as a pixel-wise gain map: 
\begin{enumerate}
    \item Gaussian mask $\mathcal{G}_{mask}(x,y) \sim \mathcal{N}_{2} ( \pmb{\mu} , \pmb{\Sigma})$ fitted from filtered sensor readings, assigns more or less intensity depending on the pixel position $(x,y)$. Its two parameters $\pmb{\mu}$ and $\pmb{\Sigma}$ are further optimized together with the end-to-end ISP model.
    \item Attention-guided mask $\mathcal{A}_{mask}$ 
    using a CNN attention block, as was illustrated in Figure~\ref{fig:main_isp}. These shallow blocks have constrained capacity to ensure the Lens Shading block only corrects per-pixel gains, and thus, we maintain the interpretability of the entire pipeline.
\end{enumerate}

Both masks are in the space $\mathbb{R}^{H \times W}$. During the reverse pass, we apply both masks to the image using an element-wise multiplication (per-pixel gain), recreating the sensor's lens shading effect. To reverse this transformation or correct the LSE, we apply the LSC mask: (i) the inverse of $\mathcal{G}_{mask}$ (element-wise divide) and (ii) $\mathcal{A}^{-1}_{mask}$ estimated by the attention block in the forward pass.


\subsection{Training}
The complete pipeline is end-to-end trainable and we can use a simple $\ell_2$ distance between the training RAW image $y$ and the estimated RAW image $\hat{y}$. To ensure the complete pipeline is invertible, we add $\ell_2$ loss terms for each intermediate image and also a consistency loss in the decomposition vectors $\mathbf{w}$ of the forward and reverse encoders. For more details we refer the reader to the supplementary material, where we also provide other relevant information about the training process \eg{ GPU devices, batch sizes, network architectures}.

\section{Experimental Results}
\label{sec:exp}

Throughout this section, we provide evidence that our method can effectively learn the RGB to RAW mapping of real unknown camera ISPs, obtaining state-of-the-art RAW reconstruction performance, and also validating the robustness of our model to operate under noisy data and data frugality (\ie few-shot learning set-ups). Additionally, we conduct experiments on a downstream task, \ie RAW image denoising, in order to validate 
ISP modelling beyond RAW image reconstruction, and the effectiveness of our proposed data augmentations. 
In all our experiments, we use the reverse pass of our model (Figure~\ref{fig:main_isp}). During the denoising experiments, we use our ISP model as an on-line domain adaptation from RGB to RAW, guided by the proposed dictionary augmentations (see Section \ref{dictionaries}). We use PSNR as a metric for quantitative evaluation defining 2 variants: PSNR$_r$ for RAW reconstruction and PSNR$_d$ for denoising.

\subsection{Datasets}
\label{sec:exp:data}
\textbf{SIDD}~\cite{Abdelhamed_2018_CVPR, Abdelhamed_2019_CVPR_Workshops}. Due to the small aperture and sensor, high-resolution smartphone images have notably more noise than those from DSLRs.
This dataset provides real noisy images with their ground-truth, in both raw sensor data (raw-RGB) and sRGB color spaces. The images are captured using five different smartphone cameras under different lighting conditions, poses and ISO levels.
There are 320 ultra-high-resolution image pairs available for training (\eg $5328 \times 3000$). Validation set consist of 1280 image pairs.
\\
\textbf{MIT-Adobe FiveK dataset}~\cite{Bychkovsky2011mit5k}. We use the train-test sets proposed by InvISP \cite{xing21invertible} for the Canon EOS 5D and the Nikon D700, and the same processing using the LibRaw library to render ground-truth sRGB images from the RAW images.

\subsection{RAW Image Reconstruction}
\label{sec:exp:data-gen-comparison}

We compare our RAW image reconstruction against other state-of-the-art methods, namely: \textbf{UPI}~\cite{Brooks_2019_CVPR} a modular, invertible and differentiable ISP model. Requires parameter tuning to fit the distribution of the SIDD dataset.
\textbf{CycleISP}~\cite{Zamir_2020_CVPR} a data-driven approach for modelling camera ISP pipelines in forward and reverse directions. For generating synthetic RAW images, we use their publicly available pre-trained model, which has been fine-tuned using the SIDD dataset. \textbf{U-Net}~\cite{unet} a popular architecture that has been previously utilized to learn ISP models~\cite{Ignatov_2020_CVPR_Workshops} as a \textit{naive} baseline trained end-to-end without any other model assumptions or regularization.

\begin{table}[t]
   \caption{Quantitative RAW reconstruction results on SIDD. The reconstruction PSNR$_r$ (dB) and top/worst 25\% are shown for each baseline method. We also show quantitative RAW denoising results in terms of PSNR$_d$ to measure the impact of the synthetic data. Additionally we include the number of parameters of each ISP model (in millions).}
   \label{table-comparison}
   \begin{center}
   \adjustbox{max width=\linewidth}{%
   \begin{tabular}{l c c c c r}
   \toprule
   Method & PSNR$_r$ & Worst 25\% & Best 25\%  & PSNR$_d$ & \# Params (M)\\
   \midrule
   UPI      & 36.84 & 14.87 & \underline{57.10} & 49.30 & 0.00 \\
   CycleISP & 37.62 & 15.90 & 51.65 & \underline{49.77} & 3.14 \\ 
   U-Net     & \underline{39.84} & \underline{20.27} & 49.61 & 49.69 & 11.77 \\ 
   Ours     & \textbf{45.21} & \textbf{21.58} & \textbf{66.33} & \textbf{50.02} &  0.59\\ 
   \bottomrule
  \end{tabular}
  }
  \end{center}
\end{table}

\begin{figure}[t]
    \centering
    \setlength{\tabcolsep}{2.0pt}
    \begin{tabular}{ccccc}
    \includegraphics[width=0.19\linewidth]{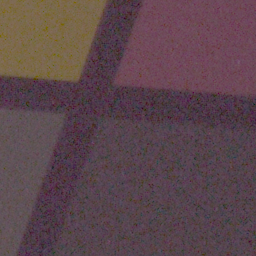} & 
    \includegraphics[width=0.19\linewidth]{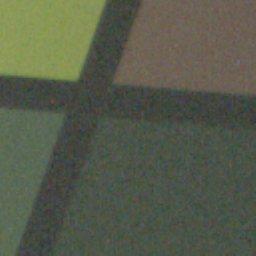} & 
    \includegraphics[width=0.19\linewidth]{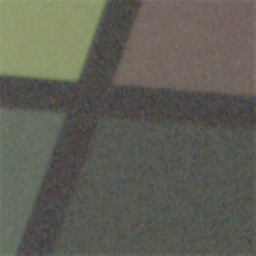} & 
    \includegraphics[width=0.19\linewidth]{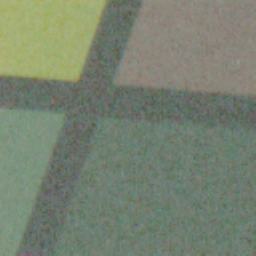} & \includegraphics[width=0.19\linewidth]{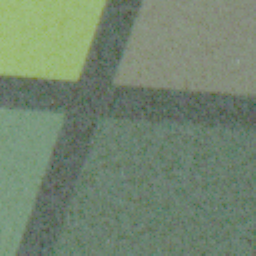}\tabularnewline
    \includegraphics[width=0.19\linewidth]{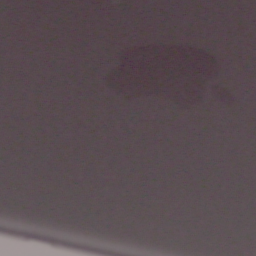} & 
    \includegraphics[width=0.19\linewidth]{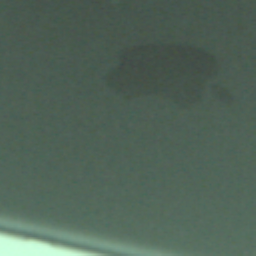} & 
    \includegraphics[width=0.19\linewidth]{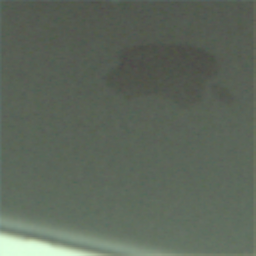} & 
    \includegraphics[width=0.19\linewidth]{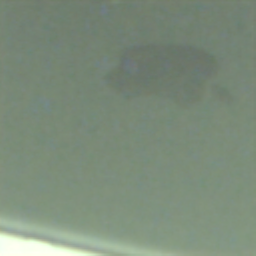} & \includegraphics[width=0.19\linewidth]{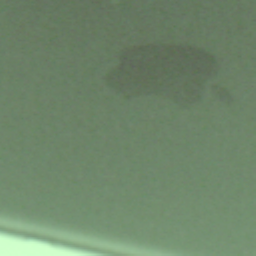}\tabularnewline
    \includegraphics[width=0.19\linewidth]{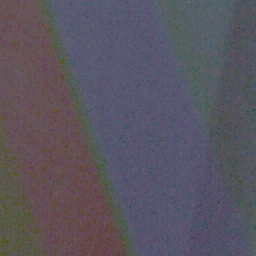} & 
    \includegraphics[width=0.19\linewidth]{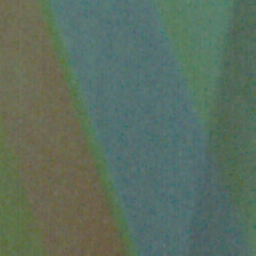} & 
    \includegraphics[width=0.19\linewidth]{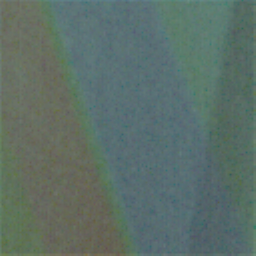} & 
    \includegraphics[width=0.19\linewidth]{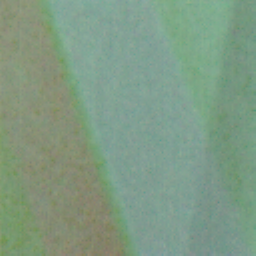} & \includegraphics[width=0.19\linewidth]{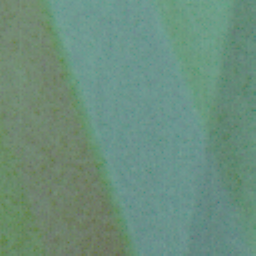}\tabularnewline
    RGB & RAW & Ours & CycleISP & UPI \tabularnewline
    \end{tabular}
    \caption{Qualitative RAW Reconstruction comparison using SIDD. Our model reconstructs better colours, tones and brightness of RAW images from different cameras.}
    \label{fig:reconst-comparison}
\end{figure}

In Table~\ref{table-comparison} we show reconstruction results in terms of PSNR$_r$ on the SIDD validation. Our model performs better than CycleISP despite being ${\sim}5{\times}$ smaller, achieving $+7.6$dB improvement, and better than U-Net despite being ${\sim}20{\times}$ smaller. We also perform better than hand-crafted methods as UPI by $+8.37$, which proves our capacity for learning camera parameters. In Figure~\ref{fig:reconst-comparison} we show a qualitative comparison of RAW reconstruction methods. 
Additionally, we aim to prove that our pipeline is invertible, by doing the \emph{cycle mapping} (sRGB to RAW and back to sRGB) our model achieves 37.82dB PSNR. More details about this experiment and qualitative results are accessible in the supplement.

Moreover, we measure the impact of the synthetic data by performing a simple denoising experiment. For a fair comparison, we use the same denoiser (U-Net) and training method as in~\citeauthor{Brooks_2019_CVPR} 
Under this setup, the only difference is the conversion from sRGB to RAW from the compared methods. We use the MIR Flickr dataset~\cite{mirflickr} as a source of sRGBs, each model transforms them into synthetic RAWs that are used for training the denoiser. These are evaluated on the SIDD Dataset. Table~\ref{table-comparison} shows the effect of our synthetic data on the denoising task, the network trained with our data achieves an improvement of $+0.7$dB PSNR$_d$ with respect to the baseline method.

\begin{figure}[t]
   \begin{center}
      \includegraphics[width=\linewidth]{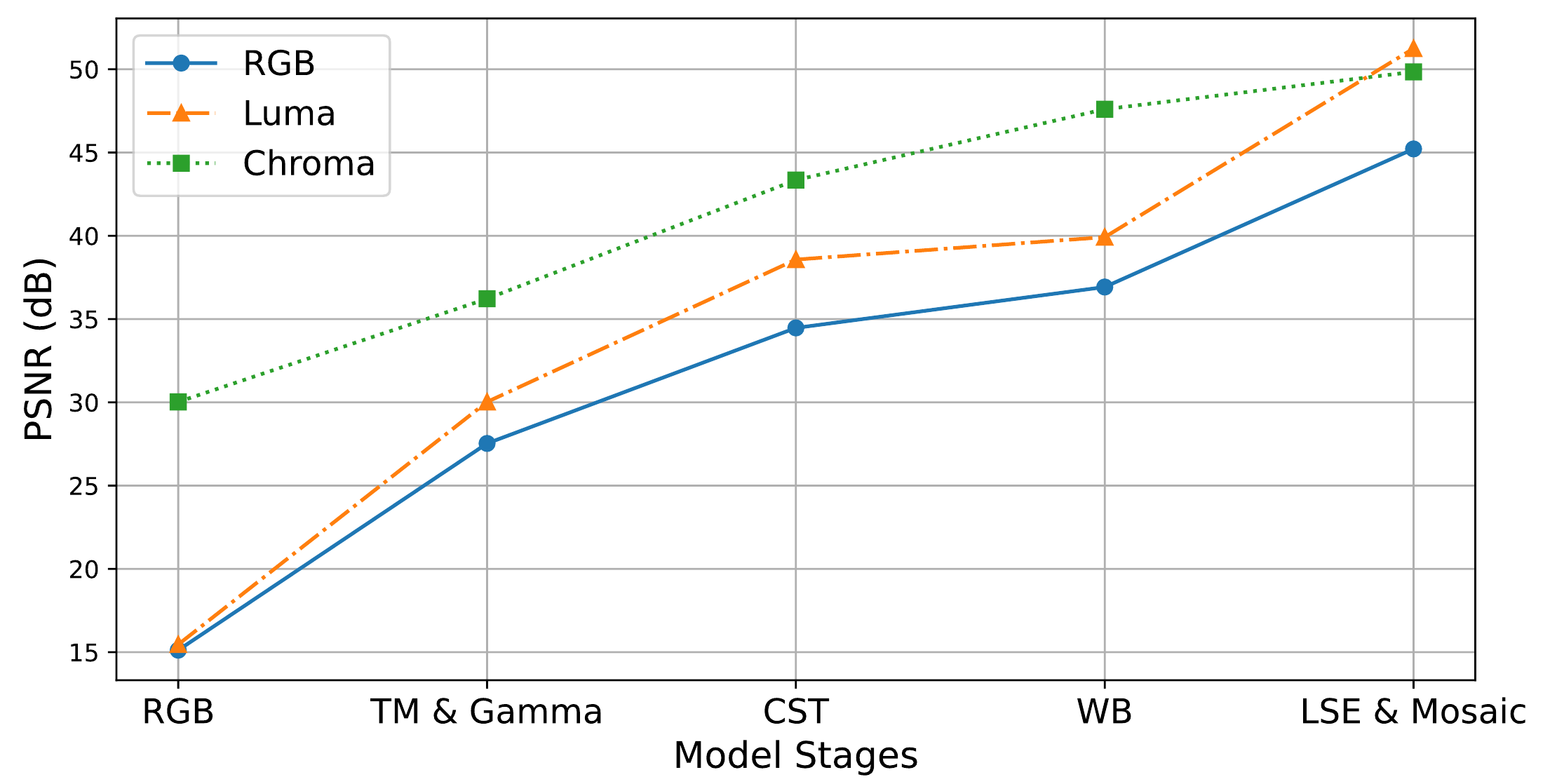}
   \end{center}
   \caption{Ablation study for the proposed techniques. We compare, in terms of PSNR, the intermediate steps $x_i$ with the original RAW image at both RGB and YUV  (luminance Y, chrominance UV) colour domains. Results show a monotonic PSNR evolution at RGB domain, meaning that after each transformation the RGB moves ''closer`` to the RAW image.}
   \label{fig:isp-ablation}
\end{figure}

Figure~\ref{fig:isp-ablation} shows the ablation of the intermediate performance of our method using the SIDD.
The monotonic PSNR evolution at the RGB domain indicates that each component in our model is contributing to improve the final reconstruction. This ablation study, together with the Table~\ref{table-comparison} quantitative results provide strong empirical evidence supporting that our pipeline and learnt parameters are realistic. The color correction and white balance (colour constancy) perform the most significant transformation in the color space, as shown by the $\ell_2$ distance reduction on the Chrominance space (the PSNR of UV components increases from 36.21dB to 43.34dB after applying our learnt CCMs, and to 47.59dB after applying our learnt WB gains). The LSE improves the Luminance space reconstruction from 39.91dB to 51.24dB.

We also test our approach using two DSLR cameras, the Canon EOS 5D and NikonD700. Using the train-test sets, loss, and patch-wise inference proposed by InvISP \cite{xing21invertible}, our method also achieves SOTA results at RAW reconstruction as we show in Table \ref{tab:nikon-table-comparison}. Note that InvISP is evaluated on the RAW with post-processed white balance provided by camera metadata, however, we do not use any information about the camera. As we have shown using the SIDD Dataset, our model is device-agnostic.

More details about the learnt parameters' distribution and qualitative comparisons are accessible in the supplement.

\begin{table}
   \caption{Quantitative RAW Reconstruction evaluation among our model and baselines proposed by Invertible-ISP~\cite{xing21invertible} using two DSLR cameras.}
   \label{tab:nikon-table-comparison}
   \begin{center}
   \adjustbox{max width=\linewidth}{%
   \begin{tabular}{l c c}
   \toprule
   Method & Nikon PSNR$_r$ & Canon PSNR$_r$\\
   \midrule
   UPI \cite{Brooks_2019_CVPR}          &  29.30 &  - \\ 
   CycleISP \cite{Zamir_2020_CVPR}    &  29.40 & 31.71 \\ 
   InvGrayscale \cite{xia18invertiblegrey} & 33.34  & 34.21 \\
   U-Net         & 38.24  & 41.52\\ 
   Invertible-ISP (w/o JPEG) & 43.29 & 45.72 \\
   Invertible-ISP (with JPEG Fourier) & \textbf{44.42} & \underline{46.78} \\
   Ours         & \underline{43.62} & \textbf{50.08} \\
   \bottomrule
  \end{tabular}
  }
  \end{center}
\end{table}

\subsection{Few-shot and Unsupervised Denoising}
\label{sec:exp-setup}
We aim to prove further the benefits of our approach and applications on downstream low-level vision tasks. In the next experiments, we use DHDN~\cite{Park_2019_CVPR_Workshops} as our denoiser. We sample shot/read noise factors as~\citeauthor{Zamir_2020_CVPR}, as such, we can inject realistic noise into the images. The mean PSNR of the noisy-clean pairs we use for training is 41.18 dB.
Our ISP model is always validated on the SIDD validation data using PSNR$_r$, our denoising experiments are validated in the same way and are reported using PSNR$_d$. We run two different experiments:\\

\textbf{Few-shot experiments:} In these experiments, we start with all available 320 sRGB-RAW \textit{clean-noisy} pairs for training our ISP model as explained in Section~\ref{sec:our-method}. In Table~\ref{sidd-benchmark} we denote the baseline method without augmentations as ``DHDN'', and the method with our ISP as on-line augmentations as ``Ours''. We can appreciate the benefit of using our approach to generate extra synthetic data, +0.46 dB improvement, and overall SOTA results. We explore how decreasing the amount of clean data available for training affects RAW reconstruction performance, and thus, RAW denoising. We experiment on three few data regimes: 184, 56, and 5 pairs available for training. Table~\ref{sidd-benchmark} shows that our denoiser trained with few-data (Ours-f) but using our augmentations, can achieve similar performance to the baseline trained with all the data.


\textbf{Unsupervised experiments: }For the last two few-shot regimes (56 and 5 pairs), we do an additional experiment where clean ground-truth denoising labels are not available. In these cases, we only use sRGB-RAW noisy pairs for training the ISP and the denoiser networks. We convert sRGBs into noisy RAWs using our augmentation strategy, and we add extra noise to our already noisy signals in order to have pseudo \textit{ground-truth} pairs~\cite{zero-shot-den}. 
Our model learns how to reconstruct RAW data even if trained with few noisy data. This ablation study and denoising results for the few-shot and unsupervised scenarios are available in the supplement. As we show in Table~\ref{sidd-benchmark}, our denoiser (Ours-u) never saw ground-truth clean images, yet when trained using our method achieved better results than models trained on real data such as DnCNN ($+6.6$ dB).


\begin{figure}[t]
    \centering
    \setlength{\tabcolsep}{2.0pt}
    \begin{tabular}{cccc}
    \includegraphics[width=0.25\linewidth]{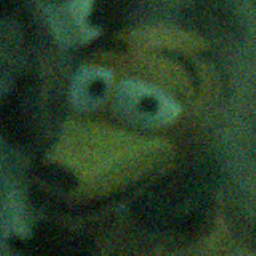} & 
    \includegraphics[width=0.25\linewidth]{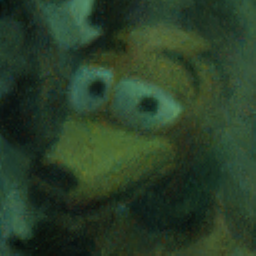} &
    \includegraphics[width=0.25\linewidth]{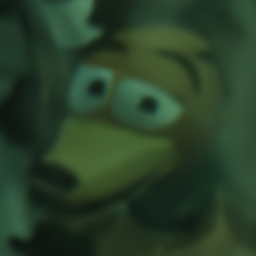} & 
    \includegraphics[width=0.25\linewidth]{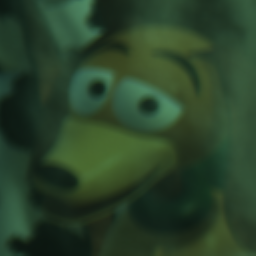} \tabularnewline
    \includegraphics[width=0.25\linewidth]{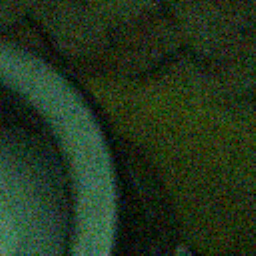} &
    \includegraphics[width=0.25\linewidth]{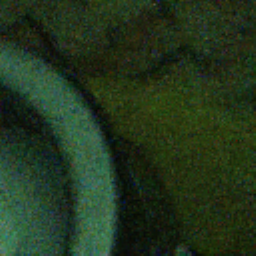} &
    \includegraphics[width=0.25\linewidth]{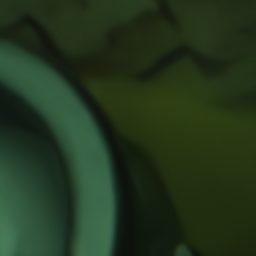} & 
    \includegraphics[width=0.25\linewidth]{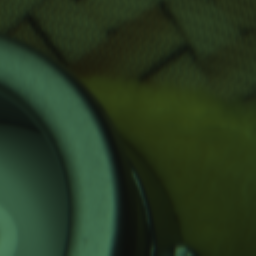}\tabularnewline
    \includegraphics[width=0.25\linewidth]{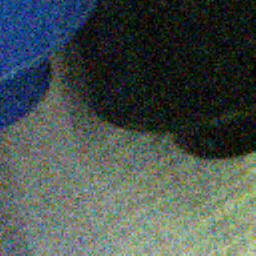} &
    \includegraphics[width=0.25\linewidth]{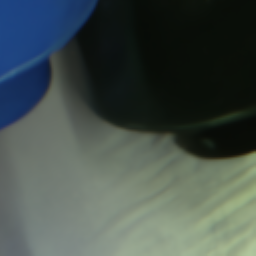} &
    \includegraphics[width=0.25\linewidth]{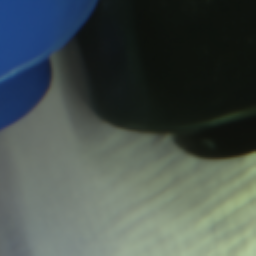} & 
    \includegraphics[width=0.25\linewidth]{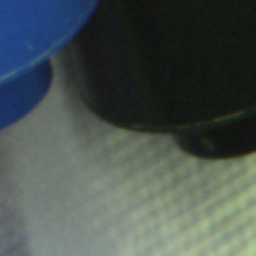}\tabularnewline
    Noisy & CycleISP & Ours & Clean \tabularnewline
    \end{tabular}
    \caption{Qualitative RAW Denoising samples. Our model removes noise while keeping textures and details. More comparisons can be found in the supplementary material.
    }
    \label{fig:den-results}
\end{figure}


\textbf{SIDD Dataset: } We report our denoising results in Table~\ref{sidd-benchmark}, 
and compare with existing state-of-the-art methods on the RAW data.
We follow a standard self-ensemble technique: four flipped and rotated versions of the same image are averaged. We use CycleISP Denoiser \cite{Zamir_2020_CVPR} publicly available weights trained on 1 million images. As shown in Table~\ref{sidd-benchmark}, our models trained on our synthetic data perform better than previous state-of-the-art despite being trained with only 320 images, and are competitive even under \textit{Few-shot} and unsupervised conditions. We conjecture that this improvement owes primarily to the various and realistic synthetic data that our method is able to leverage.

We also test our denoiser on the SIDD+ Dataset~\cite{Abdelhamed_2020_CVPR_Workshops} to show its generalization capability. Our model generalizes to new scenes and conditions, \ie we improve $1.31$ dB over CycleISP. We provide these quantitative results in the supplement.
\\

Although we have made an exhaustive study focused on Denoising, other low-level tasks \eg{ RAW data compression, Image retouching, HDR} \cite{xing21invertible} can benefit from our approach. We show a potential HDR application in the supplementary material.

\begin{table}

    \caption{RAW denoising results on the SIDD Dataset. Few-shot and unsupervised variants of our method are denoted as ``Ours-f'' and ``Ours-u'' respectively.}
    
   \begin{center}
   \begin{adjustbox}{max width=\linewidth}
   \begin{tabular}{l c c}
   \toprule
   Method & PSNR \begin{math}\uparrow \end{math} & SSIM \begin{math}\uparrow \end{math} \\
   \midrule
   Noisy & 37.18 & 0.850 \\
   GLIDE \cite{glide} & 41.87 & 0.949 \\
   TNRD \cite{tnrd} & 42.77 & 0.945 \\
   KSVD \cite{ksvd} & 43.26 & 0.969 \\
   DnCNN \cite{dncnn} & 43.30 & 0.965 \\
   NLM \cite{nlm}& 44.06 & 0.971 \\
   WNNM \cite{wnnm} & 44.85 & 0.975 \\
   BM3D \cite{bm3d} & 45.52 & 0.980 \\
   Ours-u  & 49.90 & 0.982 \\
   DHDN \cite{Park_2019_CVPR_Workshops}  & 52.02 & \underline{0.988} \\
   Ours-f  & 52.05 & 0.986 \\
   CycleISP \cite{Zamir_2020_CVPR} & \underline{52.38} & \textbf{0.990} \\
   Ours & \textbf{52.48} & \textbf{0.990} \\
   \bottomrule
   \end{tabular}
   \end{adjustbox}
   \end{center}
   \label{sidd-benchmark}
\end{table}

\subsection{Limitations}

Learning an approach to approximate inverse functions of real-world ISPs is not a trivial task for the following reasons: (i) The \textbf{quantization} of the 14-bit RAW image to the 8-bit RGB image lead to inevitable information lost. We estimate this error to be $0.0022$ RMSE for the SIDD. As previous work~\cite{Brooks_2019_CVPR, Zamir_2020_CVPR} we considered the uniformly distributed quantization error to be negligible when compared to other aspects on the RAW reconstruction problem (e.g. colour shift, brightness shift). (ii) For modern camera ISP, the operations and their parameters are unknown. Some operations as the value \textbf{clipping} can not be accurately inverted, and we have observed that the method can potentially degrade when large portions of the RGB image are close to overexposure. This is however not a common phenomena, most of the images in the datasets are properly exposed, and thus, the impact on performance is quite limited. (iii) We endeavour to 
model real camera ISPs, currently via 
six 
modules, this naturally limits performance. 
Learning to model and invert 
additional modules 
(\eg color enhancement, deblurring), 
will increase modelling power. 

\section{Conclusion}
\label{sec:conclusion}

In this paper we have proposed a novel modular, interpretable and learnable hybrid ISP model, which combines the best of both model-based and end-to-end learnable approaches. Our model performs a reversible sRGB to RAW domain transformation while learning internal ISP parameters, even under extreme low data regimes or noise. The approach recovers high quality RAW data and improves previous synthetic RAW reconstruction methods. By learning how to reverse a camera sensor and generate realistic synthetic RAW images, we can improve in downstream low-level tasks, achieving state-of-the-art performance on real camera denoising benchmarks, even with an extremely small amount of training data.



\bibliography{egbib}

\end{document}